\documentstyle[12pt]{article}

\begin{document}

\title{Two-step Liquid Drop Model for Binary, Metal-rich Clusters}
\author{F. DESPA$^{*}$ \\
{\it Laboratorium voor Vaste-Stoffysica en Magnetisme, }\\
{\it K. U. Leuven,} {\it Celestijnenlaan 200D, }\\
{\it B-3001 Leuven, Belgium}}
\maketitle
\date{}

\begin{abstract}
It is shown that differences observed between the ionization potentials of
the molecular-doped metallic clusters and those corresponding to the bare
metallic ones can be explained by a two-step approach of the classical
Liquid Drop Model. This approach takes into account the distinct physical
properties of the interface between the molecular core and the metallic
shell. Also, it is shown that the presence of the molecular core may act in
the determination of the predominant channel of the coulombic fission.

\end{abstract}

Much of the recent interest in binary clusters has focused on the stability
of atomic micro-objects consisting of metal atoms surrounding an
electronegative impurity \cite
{wu1,schleyer1,kudo1,rehm,marsden,yeretzian,jones,bonacic}. There are mainly
two interesting features exhibited in the behavior of the binary clusters.
Most striking is the appearance of non-stoichiometric structures, i.e. the
octet rule of the chemical bonding is apparently broken, in the size range
of small clusters (a few metal atoms surrounding the electronegative
element) \cite{wu1,schleyer1,kudo1,rehm,marsden,yeretzian,jones}. Such
systems are usually called hypervalent molecules. Nevertheless, by
increasing progressively the number of metal atoms, the transition to
metallicity appears as expected. It has been well demonstrated that the
passage towards the metallic state is accompanied by a structural transition 
\cite
{limberger1,limberger,rajagopal,weis,brechignac2,labastie,antoine,ZPhD,CPL,JCP}%
. Accordingly, in the size range of metal-rich clusters, a segregation shows
up between a stoichiometric ionically bound part and the excess metallic
component.

It has been suggested that the molecular core segregated around the impurity
acts to confine the valence electrons into a higher density \cite{limberger}%
. This is a quantum size effect which results in a perturbation of the
electronic density of states by a cluster-size dependent amount. Usually,
this effect can be observed in the ionization potentials \cite
{ZPhD,CPL,JCP,EPJD}. For example, the ionization potentials of the lithium
monoxide cluster $\left( Li_{n}O\right) $ are systematically larger $\left(
0.1-0.15\;eV\right) $ than the values of the bare lithium clusters $\left(
Li_{n}\right) $ in the size range $2\leq n\leq 70$ \cite{EPJD}. Although the
difference is much smaller, the same trend is also observed for lithium
monocarbide clusters $\left( Li_{n}C\right) $ \cite{EPJD}. In the context of
the quantum size effect, observed differences in the values of the size
dependent ionization potentials of $Li_{n}O$ and $Li_{n}C$ clusters in
comparison with bare $Li_{n}$ clusters have been interpreted and argued for
in terms of distinct internal contributions (the binding energy and the
surface potential barrier) to the work function \cite{EPJD}. A cluster-size
dependent work function has been derived for doped clusters which is
sensitive both to the electronegative character of the impurity and to the
size of the molecular core surrounding the impurity.

The quantum size effect as we briefly introduced above can be observed, e.g.
for clusters having spherical shapes, as long as the radial dimensions of
the molecular and metallic parts are comparable. When the metallic part is
much greater, the presence of the electronegative impurity is fully screened
by the valence electron gas and the metallic characteristics of doped
metallic clusters tend to become similar with those exhibited by bare
metallic ones. This has been proven in our previous paper \cite{EPJD}.

Nevertheless, the presence of the electronegative element with its
surrounding molecular part may affect the behavior of the binary clusters in
physical processes which can be explained without involving necessarily
their quantum aspects. For instance, one can make use of a classical Liquid
Drop Model $\left( LDM\right) $ which, generally, allows an accurate
computation of the ionization potentials and, particularly, offers a
suitable description of the fragmentation processes due to the coulombic
fission \cite{heer,brack,seidl,despa1}. Because in the present case one must
account for distinct properties of the molecular core, the $LDM$ employed in
the following is a two-step approach. At the first instant, one can
speculate on the fact that the molecular part segregated around the impurity
acts itself as an entity by inducing a tension at the interface with the
metallic outer shell. Therefore, the molecular core may behave differently
under shape deformations. This fact certainly changes the energy balance
between the charged (deformed) state of the doped cluster and that
corresponding to the bare one. This difference can be observed in the
ionization potentials \cite{ZPhD,CPL,JCP,EPJD}. On the other hand, the
molecular core can equally be seen as a mixing part which gives rise to a
configurational entropy contribution. The presence of the entropic term in
the total free energy of the system may affect the determination of the
predominant fission channel for charged (deformed) binary clusters. The
behavior of the molecular-doped metallic cluster associated with these two
particular aspects is the focus of this paper.

\medskip

The segregation of the molecular inner core changes locally the interatomic
potential and chemical bond lengths. The change of the potential energy
surface rises restoring forces acting against formation of the curved
interface between metallic and molecular parts. The fact can be expressed in
terms of interface tension \cite{henderson} 
\begin{equation}
\sigma _{m}\left( R\right) \cong \sigma _{m}\left( \infty \right) \left( 1-%
\frac{2\delta _{m}}{R}\right) \;\;.  \label{one}
\end{equation}
Here, $\left( \frac{1}{R}\right) $ is the local curvature, $\sigma
_{m}\left( \infty \right) $ is the interface tension corresponding to the
planar interface and $\delta _{m}$ is the well-defined microscopic length
(Tolman's length) \cite{henderson} depending on the characteristic length of
the interatomic potential inside the molecular part. In the same fashion,
the tension at the cluster surface can be defined as 
\begin{equation}
\sigma _{M}\left( R_{0}\right) \cong \sigma _{M}\left( \infty \right) \left(
1-\frac{2\delta _{M}}{R_{0}}\right) \;\;.  \label{two}
\end{equation}
All the parameters in above are defined with respect to the properties of
the metallic part. $\left( \frac{1}{R_{0}}\right) $ describes now the local
curvature of the outer surface of the metallic part.

The radius of the entire cluster $R_{0}$ can be approximated by 
\begin{equation}
R_{0}^{3}=R^{3}+\left( n-p-1\right) r_{s}^{3}\;\;,  \label{three}
\end{equation}
where $R$ is the radius of the molecular core, $n-1$ is the number of the
metal atoms in the cluster (we considered that one of the initial number $n$
of metal atoms has been replaced by the electronegative impurity), $p$ is
the number of host metal atoms localized by the electronegative impurity and 
$r_{s}$ is the inter-electron space (the Wigner-Seitz radii). By specifying
the atomic radius of the electronegative impurity $\left( r_{o}\right) $,
the size of the molecular core can also be set to 
\begin{equation}
R^{3}=r_{o}^{3}+pr_{s}^{3}\;\;\;,  \label{four}
\end{equation}

By looking at eq. $\left( 1\right) ,$ one can see that any modification of
the local curvature $\left( \frac{1}{R}\right) $ of the molecular core
changes the interface tension. We can ask now the question if the energy
balance between the initial and final cluster states during the physical
processes which involve shape deformations maintains the fingerprint of the
molecular core. For example, the cluster shape undergoes deformations during
the ionization process equally due to the electrifying phenomenon \cite{heer}
and thermal perturbation \cite{brechignac}. Certainly, the charged state of
a metallic cluster is a deformed one. The deformation is presumably over the
entire system, which means that the molecular core is also affected by this
change of the cluster state. It will be shown in the following that the
information about the molecular core does not cancel when considering energy
differences between the charged (deformed) and neutral states.

Certainly, the spherical approximation, usually invoked within the jellium
model for clusters \cite{brack}, becomes less useful for charged clusters
whose shapes deform due to the balance between electric and surface forces.%
\cite{heer} Complementary to the jellium model is then the Liquid Drop Model 
\cite{heer,brack}. Accordingly, when a spherical cluster of radius $R_{0},$
is deformed (conserving the volume) toward an ellipsoidal shape, its area
changes to 
\begin{equation}
A_{M}\left( \beta ^{2}\right) \cong 4\pi R_{0}^{2}\left( 1+\frac{2}{5}\beta
^{2}\right) \;\;\;,  \label{five}
\end{equation}
upon the second-order approximation in powers of the deformation parameter $%
\beta $. In this the simplest deformation, the surface energy is therefore 
\begin{equation}
E_{M}\left( \beta ^{2}\right) \cong 4\pi \sigma _{M}\left( R_{0}\right)
R_{0}^{2}\left( 1+\frac{2}{5}\beta ^{2}\right) \;\;\;.  \label{six}
\end{equation}
Also, in the simplest manner, when the cluster under consideration is
electrified, it has an electrostatic energy 
\begin{equation}
E_{C}\left( \beta ^{2}\right) =\alpha \left( r_{s}\right) \frac{e^{2}}{4\pi
\varepsilon _{0}R_{0}}\left( 1-\frac{1}{5}\beta ^{2}\right) \;\;\;.
\label{seven}
\end{equation}
where $\alpha \left( r_{s}\right) $ is a dimensionless coefficient \cite
{despa}.

As we can see, the surface energy is a minimum when the cluster is
spherical: The surface energy disfavors deformation. On the other hand, the
electrostatic energy is a maximum when the cluster is undeformed: It favors
the deformation. Consequently, the relative strengths of the electric and
surface forces will determine the shape of a charged cluster which is rather
deformed.

As a consequence of the cluster deformation, we can expect that the
molecular core undergoes deformations, too. The degree of deformation of the
molecular core depends on the strength of the molecular bonds. In the
following, we will assume that the change from the sphericity of the
molecular inner part can be measured by a deformation parameter $\gamma $,
in the same manner as for the entire system 
\begin{equation}
A_{m}\left( \gamma ^{2}\right) \cong 4\pi R^{2}\left( 1+\frac{2}{5}\gamma
^{2}\right) \;\;\;.  \label{eight}
\end{equation}
Consequently, the energy due to the interface between the ''molecular'' and
''metallic'' parts of the binary clusters becomes 
\begin{equation}
E_{m}\left( \gamma ^{2}\right) \cong 4\pi \sigma _{m}\left( R\right)
R^{2}\left( 1+\frac{2}{5}\gamma ^{2}\right) \;\;.  \label{nine}
\end{equation}
At this point one may conclude that the charged state of a cluster is a
deformed one, the deformation giving rise, in the present case, to changes
of the surface/interface energies.

The total energy of a cluster with $n$ atoms and $z$ excess electrons is
given by 
\begin{eqnarray}
E\left( n,z\right) &=&-z\Delta V+\alpha \left( r_{s}\right) \frac{\left(
ze\right) ^{2}}{4\pi \varepsilon _{0}r_{s}n^{1/3}}\left( 1-\frac{1}{5}\beta
^{2}\right) +\left( n+z\right) e_{b}+  \label{ten} \\
&&+4\pi \sigma _{M}\left( R_{0}\right) r_{s}^{2}n^{2/3}\left( 1+\frac{2}{5}%
\beta ^{2}\right) +\;...\;\;,  \nonumber
\end{eqnarray}
where $\Delta V$ is the outer part of the Coulomb barrier at the cluster
surface \cite{EPJD} and $e_{b}$ is the bulk energy. For a doped cluster, the
interface energy term given by $\left( 9\right) $ should be added to the
above expression and the size of the cluster must be replaced according to
eq. $\left( 3\right) $. The ionization potential is usually defined as a
difference in energy between the charged state $\left( z=-1\right) $ 
\begin{eqnarray}
E\left( n,-1\right) &=&WF+\alpha \left( r_{s}\right) \frac{e^{2}}{4\pi
\varepsilon _{0}r_{s}n^{1/3}}\left( 1-\frac{1}{5}\beta ^{2}\right) +ne_{b}+
\label{eleven} \\
&&+4\pi \sigma _{M}\left( R_{0}\right) r_{s}^{2}n^{2/3}\left( 1+\frac{2}{5}%
\beta ^{2}\right) +\;...\;\;,  \nonumber
\end{eqnarray}
where $WF=\Delta V-e_{b}$, and the neutral state of a system $\left(
z=0\right) $%
\begin{equation}
E\left( n,0\right) =ne_{b}+4\pi \sigma _{M}\left( R_{0}\right)
r_{s}^{2}n^{2/3}+\;...\;\;,  \label{twelve}
\end{equation}
where the deformation of the cluster has been disregarded. (We assume that
the neutral state of the cluster is spheric.)

We are now in the position to write the full equation for the ionization
potentials of molecular-doped metallic clusters. This equation can be
written as 
\begin{eqnarray}
IP &\cong &WF_{bin}+\alpha \left( r_{s}\right) \frac{e^{2}}{4\pi \varepsilon
_{0}\left[ r_{o}^{3}+r_{s}^{3}\left( n-1\right) \right] ^{1/3}}\left( 1-%
\frac{1}{5}\beta ^{2}\right) +  \label{thirteen} \\
&&\frac{8\pi }{5}r_{s}^{2}\left[ \beta ^{2}\sigma _{M}\left[ \frac{r_{o}^{3}%
}{r_{s}^{3}}+\left( n-1\right) \right] ^{\frac{2}{3}}+\gamma ^{2}\sigma
_{m}\left( \frac{r_{o}^{3}}{r_{s}^{3}}+p\right) ^{\frac{2}{3}}\right] \;\;\;,
\nonumber
\end{eqnarray}
where the distinct physical properties of the molecular inner have been
incorporated. The last term comprises corrections due to shape deformations
of the metallic and molecular parts in the final charged state of the
cluster. We observe that by replacing in $\left( 10\right) $ $\gamma =0,$ $%
p=0$ and $r_{o}$ by $r_{s}$ the equation for ionization potentials for bare
metallic clusters is recovered 
\begin{equation}
IP_{bare}\cong WF+\alpha \left( r_{s}\right) \frac{e^{2}}{4\pi \varepsilon
_{0}r_{s}n^{1/3}}\left( 1-\frac{1}{5}\beta ^{2}\right) +\frac{8\pi }{5}%
r_{s}^{2}\beta ^{2}\sigma _{M}n^{\frac{2}{3}}\;\;.  \label{fourteen}
\end{equation}
The deformation parameters $\beta $ is inversely proportional with the
cluster size \cite{krane}, 
\begin{equation}
\beta =\frac{2}{3}\frac{\Delta \Gamma }{R_{0}}\equiv \frac{2}{3}\frac{\Delta
\Gamma }{r_{s}n^{\frac{1}{3}}}\;\;\;,  \label{fiveteen}
\end{equation}
where $\Delta \Gamma $ is the difference between the semimajor and semiminor
axes of the ellipsoid. This cancels exactly with the cluster-size dependence
of the last terms in $\left( 14\right) $ and makes the surface contribution
to depend only on $\Delta \Gamma $. In the limit of very large clusters this
contribution becomes negligible $\left( \Delta \Gamma =0\right) $ and, in
the asymptotic limit, $IP_{bare}\rightarrow WF$. The same applies in $\left(
13\right) $ with regard both to the surface and interface contributions. As $%
\gamma $ is directly related to $\beta $, the deformation of the molecular
part during the ionization process being a consequence of the deformation
over the entire system, the influence of the molecular core will also
diminish in the limit of large clusters.

\medskip Looking at $\left( 13\right) $ and $\left( 14\right) ,$ one can see
that, apart the surface/interface contributions, the other factors which
promotes the difference between the ionization potentials of the binary,
metal-rich and bare metallic clusters are the reduction of the valence
electron number of the cluster ($n-p$ instead of $n$) and the change of the
work function (the usual work function $WF$ is replaced in $\left( 13\right) 
$ by a work function for binary clusters, that is $WF_{bin}$) \cite{EPJD}.
This change of the work function for doped clusters is a quantum size effect
due to the perturbation of the electron density. This difference disappears
with increasing cluster size, the impurity being fully screened by the
remaining free electrons for a cluster with about $100$ atoms, as it has
been proven in our previous work \cite{EPJD}.

In the following, we disregard the distinct internal contributions to the
work function of doped clusters arising from the quantum size effect ($%
WF=WF_{bin}$), and focus on the surface/interface contributions promoted by
distortions of the cluster shape. We derive the ratio between $IP$ and $%
IP_{bare}$ , which is 
\begin{equation}
\frac{IP}{IP_{bare}}=\left( \frac{r_{o}^{3}}{nr_{s}^{3}}+\frac{n-1}{n}%
\right) ^{2/3}+\frac{\gamma ^{2}}{\beta ^{2}}\frac{\sigma _{m}}{\sigma _{M}}%
\left( \frac{r_{o}^{3}}{nr_{s}^{3}}+\frac{p}{n}\right) ^{2/3}\;\;\;,
\label{sixteen}
\end{equation}
We make a numerical test for the above result and compare with the
experimental observation as reported in Ref. \cite{EPJD}. For example, in
the size domain of of large clusters $\left( n>6\right) $, the experimental $%
IP^{\prime }s$ of oxygen doped lithium clusters are systematically larger
(about $4\%$) than the values of the bare lithium clusters. By looking to
the above equation $\left( 16\right) $, we see that for $n\gg 1$, this can
can be approximated by 
\begin{equation}
\frac{IP}{IP_{bare}}\simeq 1+\frac{\gamma ^{2}}{\beta ^{2}}\frac{\sigma _{m}%
}{\sigma _{M}}\left( \frac{p}{n}\right) ^{2/3}\;  \label{seventeen}
\end{equation}
which leads to 
\begin{equation}
\frac{IP-IP_{bare}}{IP_{bare}}=\frac{\gamma ^{2}}{\beta ^{2}}\frac{\sigma
_{m}}{\sigma _{M}}\left( \frac{p}{n}\right) ^{2/3}\;\;.  \label{eighteen}
\end{equation}
We set $p=2$ and $n=20$ (this is supposed to be the case of $\left(
Li_{2}O\right) Li_{18}$ cluster) and observe that the difference $%
IP-IP_{bare}$ of about $4\%$, as reported in Ref. \cite{EPJD}, can be
obtained if either the tension is higher at the surface than at the
interface $\left( \sigma _{M}\simeq 5\sigma _{m}\right) $ at $\gamma =\beta $%
, or, equivalently, $\frac{\gamma ^{2}}{\beta ^{2}}\leq 0.04$ at $\sigma
_{M}\simeq \sigma _{m}$. The latter means that the molecular core is less
affected by the distortion of the entire cluster. The molecular core opposes
to the external action exerted by the metallic outer shell. Indeed, both
effects can concur to set the above difference.

At small cluster sizes, when the charged clusters are rather deformed, the
interface contribution in $\left( 9\right) $ can play an important role.
This has a material dependence by $\sigma _{m}\left( R\right) $. There is a
sizeable body of work on aspects of the statistical mechanical theory of
interfacial phenomena \cite{rowlinson}. Anyway, it is difficult to say what
results tell us about the physical properties of a real interface. The
concept of a bare surface/interface tension is controversial, since
correlation effects for the restoring force contribute also to the specific
thermodynamical potential. Looking at $\left( 1\right) ,$ we can see that
the curvature affects the surface tension. The effect comes to light at full
extent for microscopic drops whose radii approach the range of
intermolecular forces. The Tolman length $\delta $ entering $\left( 1\right) 
$ has negative values \cite{henderson} 
\[
\delta \cong -0.39\xi \;\;\;, 
\]
which depends on the characteristic lengths of the intermolecular
potentials, $\xi $. Obviously, we can imagine that different properties of
the binary clusters can be explained by taking the particular chemical
structure of the molecular part into account. Some geometrical structures of
the molecular part can fetter the ellipsoidal deformation of the entire
system or, by contrast, others can favor it. So that, the geometrical
structure of the molecular part can play a role in moderating the distortion
tendency of the charged clusters.

\medskip

Another aspect related to the presence of the molecular inner core is the
behavior of the binary clusters during the fragmentation processes by
coulombic fission. It was shown that, generally, the binary fission occurs
predominantly with asymmetrical character \cite{brechignac}. The discrete
nature of the cluster constituents become important in this case, since the
same fragments may consist of different atoms. In order to see how the
molecular core can affect the determination of the predominant fission
channel, we describe the fission process 
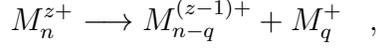
\[
M_{n}^{z+}\longrightarrow M_{n-q}^{\left( z-1\right) +}+M_{q}^{+}\;\;\;, 
\]
as a matter of minimizing the free energy \cite{brechignac,brechignac3} 
\begin{equation}
F=\Sigma +C-k_{B}TS\;\;,  \label{nineteen}
\end{equation}
where 
\begin{eqnarray}
\Sigma &=&a_{f_{q}}q^{2/3}+a_{f_{n-q}}\left( n-q\right)
^{2/3}-a_{n}n^{2/3}\;\;,  \label{twenty} \\
C &=&-\frac{e^{2}}{4\pi \varepsilon _{0}r_{s}}\left[
c_{n}n^{-1/3}-c_{f_{n-q}}\left( n-q\right) ^{-1/3}-c_{f_{q}}q^{-1/3}\right]
\;\;,  \nonumber
\end{eqnarray}
are the surface and coulombic energies. $a_{f_{i}}$ $\left( i=q,n-q\right) $
and $a_{n}$ stand for the surface energies per atom for the fission
fragments and parent sample. $c$ is the coulombic energy per atom and the
subscripts have the same meaning as above. The entropy $S$ is given by 
\begin{eqnarray}
\frac{S}{k_{B}} &=&\ln \frac{n!}{\left( n-q\right) !\;q!}  \label{twentyone}
\\
&\cong &-n\left[ \frac{q}{n}\ln \frac{q}{n}+\left( 1-\frac{q}{n}\right) \ln
\left( 1-\frac{q}{n}\right) \right] \;\;,  \nonumber
\end{eqnarray}
and shows that the same fragments may consist of different atoms. We notice
that in the extreme case when one of the fission product, let say $q$, is
just the molecular core, the surface $\left( a_{f_{q}}\right) $ and
coulombic $\left( c_{f_{q}}\right) $ energies, which control the fission
process, must be properly replaced in $\left( 20\right) $. The surface term
corresponding to the fragment $q$ can be derived starting from $\left(
1\right) $. The estimation of the coulomb energy of the fragment $q$ must
account for the appropriate character of the chemical bond in the molecular
core. To compare with the trivial case of bare metallic clusters, we make
the terms $a_{f_{i}}$ and $a_{p}$ equal each other $\left( a_{f_{i}}\equiv
a_{p}=a\right) $, and replace $c_{f_{q}}=\frac{3}{8}$ $c_{f_{n-q}}=\frac{7}{4%
}$ and $c_{p}=\frac{33}{8}$ \cite{brechignac3}. We observe that the
predominant fission channel for binary clusters is affected by the physical
properties of the molecular core. We can also see that if the molecular core
is more stable than the rest of the metallic cluster then, the size of the
molecular core acts as a lower bound for the characteristic mass of the
fission product $\left( q>p+1\right) $ \cite{brechignac2}.

\bigskip

Finally, the present study on binary, metal-rich clusters based on a
two-step $LDM$ approach allowed us to point out their different behavior in
comparison with the corresponding bare metallic species. This behavior has a
general feature for metal-rich clusters, in agreement with the segregation
evidence of the molecular part: The molecular core acts itself as an entity
by inducing a tension at the interface with the metallic outer shell and as
a mixing part by giving rise to a configurational entropy contribution. The
former depends on the chemical character of the binding and its energetic
contribution is in direct proportion with the interface area. The latter
comes to light at full extent for rather high temperatures. The effect of
the former leads to an energetic balance in the final (charged) state of the
cluster wherefrom the difference between the ionization potentials of the
doped clusters and corresponding pure clusters shows up. The latter affects
the determination of the predominant fission channel of the coulombic
fission. The size of the molecular core can act as a lower bound for the
characteristic mass of the fission product.

Concluding, one can say that, although the metallic characteristics
definitely predominate in the range of large cluster sizes, the
electronegative impurity still affects (not dramatically, indeed) their
thermodynamical properties. Thus, the molecular core of the binary,
metal-rich cluster has a response to external actions, this resulting in a
change of the total energy of the system.

\bigskip

\bigskip

$^{*}$Present address: Department of Chemistry, The University of Chicago,
5735 S. Ellis Avenue, Chicago, IL 60637, e-mail: fdespa@midway.uchicago.edu

{\bf Acknowledgments}

This project has financially been supported by the Fund for Scientific
Research - Flanders (Belgium) (F.W.O.). I thank R.E. Silverans and P.
Lievens for their kind hospitality during my stay in Leuven.


\begin{thebibliography}{99}
\bibitem{despa}  The dimensionless coefficient $\alpha $ entering the
electrostatic energy is often used as a fitting parameter for the
experimental curves. In a classical approach, $\alpha $ is given by $\alpha
\left( r_{s}\right) \simeq \frac{1}{2}\left( 1+\frac{z}{9}\frac{r_{s}^{3}}{%
R_{0}^{3}}+\frac{z^{2}}{27}\frac{r_{s}^{6}}{R_{0}^{6}}+...\right) $, with $z$
standing for the number of the elctric charges removed from the cluster. For
more details see the papers by F. Despa, Z. Phys. {\bf D 37} 347 (1996) and
by M. Seidl and J.P. Perdew, Phys. Rev. {\bf B 50} 5744 (1994).
\end{thebibliography}
\end{document}